\documentclass[twocolumn,prb,showpacs]{revtex4}

\newcommand{\lcmo}{La$_4$Cu$_3$MoO$_{12}$}

\usepackage{graphicx}

\begin{document}
\title{\bf  Spin-Trimer Antiferromagnetism in La$_4$Cu$_3$MoO$_{12}$}
\author{Y. Qiu}
\email{qiuym@pha.jhu.edu}
\affiliation{Department of Physics and Astronomy, The Johns Hopkins University, Baltimore, MD 21218}
\author{C. Broholm}
\affiliation{Department of Physics and Astronomy, The Johns Hopkins University, Baltimore, MD 21218 and\\
NIST Center for Neutron Research, National
Institute of Standards and Technology, Gaithersburg, MD 20899}
\author{S. Ishiwata}
\author{M. Azuma}
\author{M. Takano}
\affiliation{Institute for Chemical Research, Kyoto University, Uji, Kyoto-fu 611-0011, Japan}
\author{R. Bewley} 
\affiliation{ISIS Facility, Rutherford Appleton Laboratory, Chilton, Didcot OX11 0QX, United Kingdom }
\author{W. J. L. Buyers}
\affiliation{National Research Council, Chalk River Laboratories, Chalk River, Ont, K0J 1J0, Canada}
\date{\today}
\begin{abstract}
\lcmo\ is a cluster antiferromagnet where copper spin-1/2 form a network of strongly coupled spin-trimers. The magnetic properties of this material have been examined using magnetic neutron scattering. At low temperatures, excitations from the ground state are observed at 7.5(3) meV and 132.5(5) meV. An additional peak in the neutron scattering spectrum, which appears at 125.0(5) meV on heating is ascribed to a transition between excited states. The wave-vector and temperature-dependence of the inelastic magnetic scattering cross section is consistent with intra-trimer transitions. Magnetic neutron diffraction reveals antiferromagnetic order below T$_N$=2.6 K that doubles the unit cell along the {\bf a} direction. The ordered magnetic structure is described as inter-trimer order where spin correlations within trimers are controlled by the strong intra-trimer interactions. Combining the information derived from elastic and inelastic magnetic neutron scattering with group theoretical analysis, a consistent set of intra-trimer interactions and ordered magnetic structures are derived. The experiment provides a simple worked example of magnetism associated with inter-atomic composite degrees of freedom in the extreme quantum limit.
\end{abstract}

\maketitle

\section{INTRODUCTION}

Geometrically frustrated magnets are distinguished by an anomalous cooperative paramagnetic phase extending to temperatures well below characteristic microscopic energy scales.\cite{Ramirez} Fluctuations in this phase are strong and non-trivial in that they satisfy certain local constraints. In some cases, further cooling fails to produce a phase transition and the low temperature state is quantum disordered\cite{phcc,cuhpcl}. In other cases, however, thermal or quantum fluctuations, magneto-elastic coupling, impurities, or sub-leading exchange interactions yield a finite value of $\langle{\bf S}\rangle$ at sufficiently low temperatures\cite{zncro,scgo,ymoo,shl}.

\begin{figure}
\includegraphics[width=3.2in]{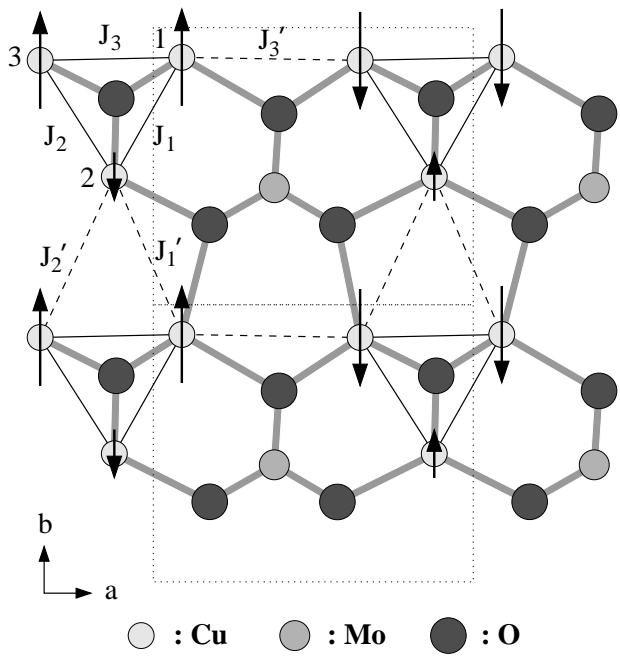}
\caption{\label{fig1}Cu$_3$MoO$_4$ plane of \lcmo. Actual coordinates for copper atoms are given in appendix A. The solid lines represent the triangle clusters. J$_1$, J$_2$ and J$_3$ are the intratriangle couplings; J$'$s are the weak intertriangle couplings that are assumed to be antiferromagnetic and of similar magnitude as those in the mean field analysis of Wessel and Haas.\cite{Wessel} The arrows illustrate one plane of the $\tau_2$ or $\tau_3$ magnetic structures with $\psi=0$ listed in table~\ref{tab:spin}.}
\end{figure}

To advance the understanding of geometrically frustrated magnetism, we have explored spin correlations in \lcmo, a material which affords a spectacular example of geometrical frustration in the quantum limit. \lcmo\ is a monoclinic ABO$_3$ type cuprate with space group P112$_1$/m and B-cations in 1:3 ratio.\cite{Vander_Griend}  Shown in Fig.~\ref{fig1}, the Cu$_3$MoO$_4$ layers of the material are built from Cu$_3$O triangular clusters, which we shall denote trimers. Consideration of the crystal structure and the Goodenough rules\cite{bibjbg} indicate that intra-trimer exchange interactions are orders of magnitude stronger than inter-trimer interactions.  Magnetic susceptibility data\cite{Azuma,Vander_Griend} show two distinct linear regimes for $1/\chi (T)$ versus temperature. For $T>400$ K the Curie Weiss fit yields $\mu_{eff}$=1.81  $\mu_B$ and  $\Theta_W$=$-558$ K, which is consistent with strong intra-trimer interactions. For  $T<250$ K the Curie constant decreases by a factor of 0.39 and the Weiss constant decreases to $-16$ K. This indicates a cross over to a cooperative paramagnetic phase where each trimer represents a composite spin-1/2 degree of freedom.  The small Weiss temperature in this phase indicates weak AFM inter-trimer interactions. At lower temperatures still, a maximum in $\chi (T)$ and a peak in the specific heat indicates an antiferromagnetic phase transition at $T_N=2.6$ K.\cite{Azuma}

In this paper, we present a comprehensive neutron scattering study of \lcmo. Polarized and unpolarized inelastic  neutron scattering data provide evidence for magnetic excited states 7.5 meV and 132.5 meV above the ground state. A calculation based on a simple intra-trimer Hamiltonian is presented to explain the energy levels, the wave vector dependence, and the temperature dependence of the inelastic neutron scattering cross section. Using elastic neutron diffraction we also show that \lcmo\ has long range antiferromagnetic order at low temperatures. While a unique ordered structure cannot be identified directly from the diffraction data, analysis based on ordering of composite spin-1/2 degrees of freedom on neighboring trimers yields a consistent set of intra-trimer exchange interactions and inter-trimer spin configurations. The experiments and analysis provide a comprehensive understanding of geometrically frustrated quantum magnetism in a simple model system with important analogies to more complex systems. 

\section{EXPERIMENTAL RESULTS}
A powder sample of \lcmo\ was synthesized using a previously published 
method.\cite{Azuma} Rietveld analysis of powder neutron diffraction data 
confirmed a single phase sample with space group P112$_1$/m and lattice 
parameters $a=7.9119(5)$ \AA, $b=6.8588(4)$ \AA,  $c=10.9713(3)$ \AA, and 
$\gamma=90.008(9)^\circ$ at 10K.  The elastic neutron scattering 
measurements were 
carried out on the BT2 thermal neutron triple-axis spectrometer at the NIST Center for Neutron Research. For that experiment we used a 30.2 g powder sample in a cylindrical container with a diameter of 1.6 cm. Pyrolytic Graphite (PG) crystals set for the (002) reflection were used to select 14.7 meV neutrons for diffraction.  There was a PG filter in the incident beam to suppress higher order contamination and collimations were $60'-40'-40'-200'$ through the instrument from source to detector. 

Inelastic neutron scattering measurements were performed on the HET direct geometry time-of-flight spectrometer at ISIS pulsed spallation neutron source. For that experiment we used a 80.2 g powder sample in a cylindrical container with a diameter of 3.1 cm. We used that instrument's ``sloppy'' chopper at frequencies of revolution 250 Hz and 400 Hz for incident energies of 40 meV and 160 meV respectively. 

Polarized inelastic neutron scattering measurements were performed on the C5 triple-axis spectrometer at the NRU reactor in Chalk River Laboratories in Canada. Magnetized Heusler crystals set for the polarizing (111) reflection were used as monochromator and analyzer with the latter fixed to reflect 14.6 meV neutrons. A cold sapphire filter  was placed before the monochromator to eliminate high-energy neutrons. A PG filter was placed in the scattered beam to suppress order contamination at the analyzer. An energy dependent correction was applied to the incident monitor count rate to account for incident beam $\lambda /2$ contamination. For this experiment we used a 120 g powder sample in a cylindrical container with a diameter of 2.5 cm. The collimation was $45'$ in the incident and $80'$ in the scattered beam. A Mezei flipper was placed in the scattered beam. The flipping ratio measured at 14.6 meV on a powder reflection was 24:1.

Absolute normalization turned out to be crucial to derive information about the magnetic structure. We used nuclear Bragg scattering to normalize the magnetic Bragg peak intensity.\cite{shl} The inelastic neutron scattering data was normalized by comparison to incoherent elastic count rates from a 20.14 g vanadium sample in the ISIS experiment. A correction for neutron absorption in \lcmo\ was applied in this latter case.

\subsection{Elastic Neutron Scattering}
\begin{figure}
\includegraphics[width=3.2in]{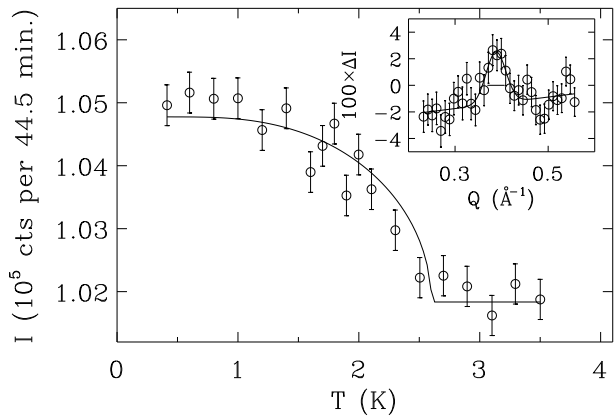}
\caption{\label{fig2}Temperature dependence of the ($\frac{1}{2}$00) magnetic Bragg peak intensity. The inset shows the wave vector dependence of the difference between elastic magnetic neutron scattering at $T=0.4$ K and $T=10$ K. The energy resolution was 1.3 meV and the wave vector resolution is indicated by the solid line in the inset.} 
\end{figure}

Fig.~\ref{fig2} shows the temperature dependence of magnetic neutron diffraction at Q=0.397 \AA$^{-1}$. The data is evidence for an antiferromagnetic phase transition at T$_N$=2.6 K $\ll\Theta_W$. The inset shows temperature difference data indicating that elastic scattering from the low temperature spin structure is in the form of a well-defined Bragg peak. To within error the location of the peak is at $\frac{1}{2}a^*$ implying that the magnetic order doubles the unit cell in the $\bf a$ direction. After due consideration of the finite instrumental resolution (solid bar in the inset), we derive a lower limit of 100 \AA~ for the magnetic correlation length. 

\subsection{Inelastic Neutron Scattering}
\begin{figure}
\includegraphics[width=3.2in]{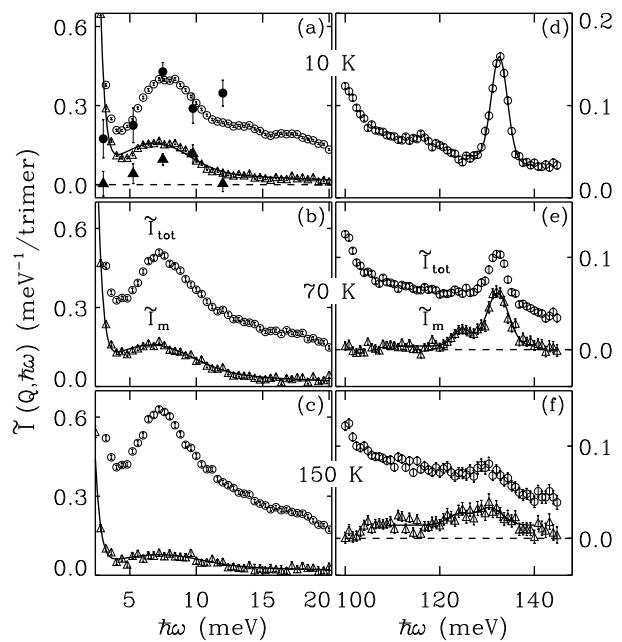}
\caption{\label{fig3}$\hbar\omega$-dependence of the normalized Q-averaged neutron scattering intensity $\tilde{I}$(Q,$\hbar\omega$) at various temeratures and in two different energy ranges. The open circles are raw experimental data. The open triangles are the phonon background subtracted data in (a)--(c), and the excess scattering above that at 10 K  for (e)--(f). The closed circles and closed triangles in (a) are total scattering and magnetic scattering intensities respectively from the polarized neutron measurement. A single scale factor was applied to the polarized data for best agreement with the unpolarized time of flight data.}
\end{figure}

Fig.~\ref{fig3} shows the energy dependence of the scattering cross section at various temperatures. Because of the wide dynamic range, we show only the interesting magnetic parts of the spectrum, low energies on the left and higher energies on the right. The right panels clearly show an excitation at 132.5 meV. As temperature increases, the peak intensity at 132.5 meV decreases and a second peak emerges at $\hbar\omega$=125 meV. A likely explanation for the lower energy peak is that it corresponds to a transition to the 132.5 meV state from an  excited state at 132.5 meV-125 meV=7.5 meV, which is populated on heating. To explore this scenario, the left panels focus on the energy range around 7.5 meV. As is apparent from the increase of intensity with temperature, there is significant phonon scattering in the lower energy range and this complicates the task of isolating the magnetic contribution to the scattering cross section. Since phonon scattering increases with temperature\cite{Lovesey} while low energy magnetic scattering generally decreases with temperature, we used high temperature data ($T=200$ K) where magnetic scattering is negligible to determine the phonon density of states. At each temperature, the appropriate thermal factor was then applied to yield the phonon contribution to inelastic neutron scattering. The  triangles in the left panels of Fig.~\ref{fig3} show the background subtracted data that indicate a residual peak in the excitation spectrum, which we tentatively associate with magnetic scattering.

To unambiguously verify the existence of a magnetic excitation at 7.5 meV, a polarized neutron scattering measurement was carried out at T=6 K and at a momentum transfer of 1.5 \AA$^{-1}$. Only magnetic scattering can produce a difference between spin flip scattering with guide fields at the sample position parallel and perpendicular to wave vector transfer.\cite{mrk} Such difference data is shown as closed triangles in Fig.~\ref{fig3}(a) and they provide immutable evidence for a magnetic excitation at 8(1) meV. Also shown is the sum of spin flip and non-spin-flip scattering as solid circles. Twice the count rate with the analyzer turned through 4 degrees  was subtracted from the summed data and a single overall scale factor was applied to all the polarized data to facilitate comparison to the time of flight data (open circles).   The excellent agreement between the two independent determinations of the magnetic contribution to inelastic scattering at 7.5 meV provides strong evidence for an intra-trimer excited state at this energy.

\begin{figure}
\includegraphics[width=3.2in]{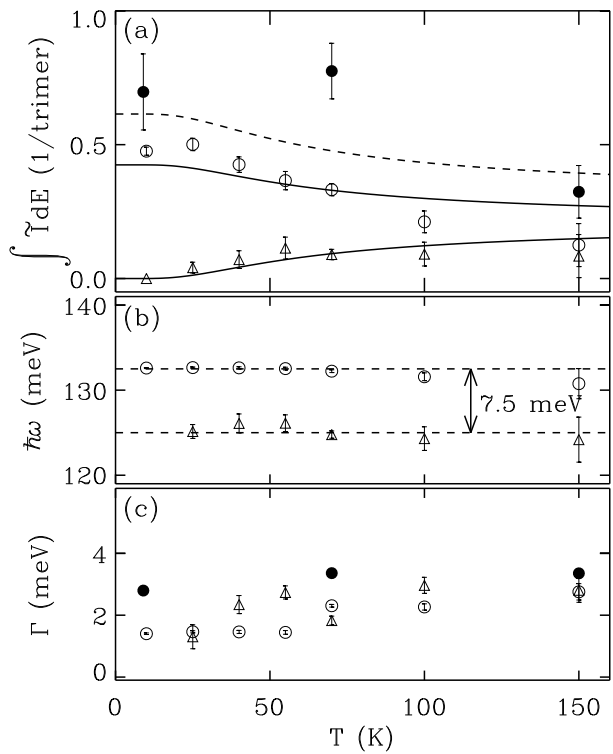}
\caption{\label{fig4}Temperature dependence of parameters characterizing high and low energy magnetic excitations in \lcmo. Frame (a) shows the integrated intensities for the 132.5 meV (open circles), the 125 meV (triangles), and the 7.5 meV (closed circles) modes. Frame (b) shows the peak positions and frame (c) shows the intrinsic half width at half maximum relaxation rate for each mode. Lines in frame (a) were calculated from the trimer model.} 
\end{figure}

From gaussian fits to all inelastic data, we derived the temperature dependence of the integrated intensities, peak positions and peak widths that are reported in Fig.~\ref{fig4}. For peak widths, the calculated energy resolution was subtracted in quadrature to produce values for the intrinsic half width at half maximum relaxation rate, $\Gamma$ or inter-trimer bandwidth. The energy levels are temperature independent to within error as expected for intra-trimer excitations. $\Gamma$ is of order the low-$T$ Curie-Weiss temperature, which is consistent with inter-trimer interactions being the main source of intra-trimer level broadening. Transitions involving the excited state doublet appear to have a greater relaxation rate than those involving other levels. Possible reasons for this include magneto-elastic effects and enhanced inter-trimer coupling for trimers occupying the 7.5 meV excited state. 

\begin{figure}
\includegraphics[width=3.2in]{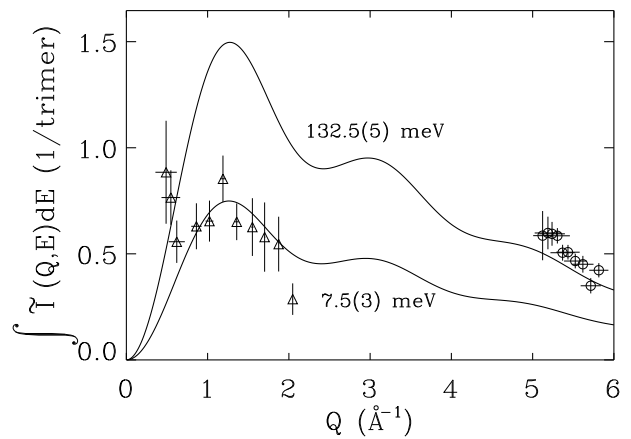}
\caption{\label{fig5} Wave-vector dependence of energy integrated intensity for the 132.5 meV (circles) and the 7.5 meV excitations (triangles) in \lcmo. Solid lines show the calculated Q-dependence of the neutron scattering cross section from Eq. (\ref{eq:B5}) and (\ref{eq:B6}) with no adjustable parameters.} 
\end{figure}

Fig.~\ref{fig5} shows the wave-vector dependence of the energy integrated intensity for the 7.5 meV and the 132.5 meV modes. Kinematical limitations prevented measurement of the high energy excitations over a significant range of wave-vector transfer and at lower energies admixture of phonon scattering complicates the analysis. Still there is evidence that the intensity of the 7.5 meV mode decreases in rough correspondence with the magnetic form factor for copper, which indicates that this excitation involves a very small cluster of copper atoms. In addition we shall see that the relative intensity of the two modes is consistent with a simple trimer exchange Hamiltonian.

\section{ANALYSIS AND DISCUSSION}
\subsection{Energy Level Scheme}
\label{energylevels}

\begin{figure}
\includegraphics[width=3.2in]{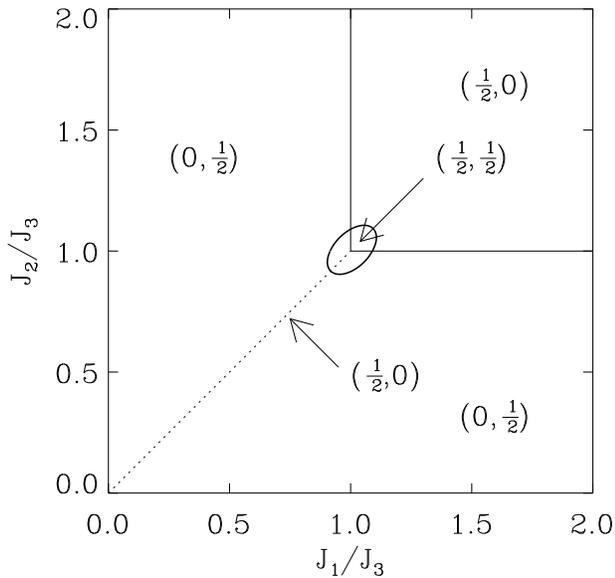}
\caption{\label{fig6}Zero-temperature magnetic phase diagram of the AFM spin-1/2 trimer square lattice with a weak intertriangle coupling J$'$=0.01J$_3$. The excitation energies observed in the present experiment imply that the ratios of exchange constants lie on the ellipse shown in the center of the figure. The observation of $(\frac{1}{2} 00)$ magnetic order also helps to constrain possible values of the exchange constants.} 
\end{figure}

From the energy level scheme we can extract detailed information about intra-trimer exchange interactions. The highest energy state for a spin-1/2 trimer with antiferromagnetic Heisenberg interactions is a quartet with total spin 3/2. If the interactions within the trimer have the symmetry of an equilateral triangle there is a fourfold degenerate ground state composed of two degenerate Kramers doublets. For lower symmetry the degeneracy is lifted and there are two low energy doublets. While they approximate the symmetry of isosceles triangles, the spin-1/2 trimers in \lcmo\ have no exact symmetry elements. In Appendix B, we calculate the energy levels and scattering cross sections for a general spin-1/2 trimer with a model Hamiltonian of the form
\begin{equation}
\label{eq:1}
H=J_1 {\bf S_1 \cdot S_2}+J_2 {\bf S_2 \cdot S_3} +J_3 {\bf S_3 \cdot S_1}~. 
\end{equation}
The splitting between the doublets is E$_{01}$=$\sqrt{J_1^2+J_2^2+J_3^2-J_1J_2-J_2J_3-J_3J_1}$ and the splitting between the ground state doublet and the quartet is E$_{02}$=$\frac{1}{2}(J_1+J_2+J_3+E_{01})$. These energies can be associated respectively with the 7.5 meV and the 132.5 meV transitions observed by inelastic neutron scattering. From this we can derive the average intra-trimer exchange constant to be $\bar{J}=(2E_{02}-E_{01})/3=85.8$ meV. The splitting between doublets yields information about the ratios between intra-trimer exchange constants. Ratios that are consistent with the 7.5 meV doublet-doublet transition lie on an ellipse in the $J_1/J_3$ versus $J_2/J_3$ plot shown in Fig.~\ref{fig6}. The ellipse is centered at the equilateral point (1,1) and the major axis lies along the isosceles $J_1=J_2$ line. The half major axis is approximately $\sqrt{9E_{01}^2/(2E_{02}^2-2E_{01}E_{02})}=0.124$ and the half minor axis is approximately $\sqrt{3E_{01}^2/(2E_{02}^2-2E_{01}E_{02})}=0.071$. In the following, we shall label intra-trimer interaction parameters that are consistent with the spectroscopic information by the counter-clockwise azimuthal angle, $\psi$, on this ellipse with $\psi=0$ corresponding to $J_1=J_2$ and $\psi=n\pi$ describing isosceles triangles. 

Wessel and Hass\cite{Wessel} recently studied the phase diagram for \lcmo\ using a model Hamiltonian including the nearest neighbor interactions within the a-b plane defined in Fig.~\ref{fig1}. This study shows that different ratios of intra-trimer interactions yield different in-plane wave vectors for long range order. The wave vectors predicted in the limit of vanishing inter-trimer interactions at $T=0$ are indicated on Fig.~\ref{fig6}. Given that $T_N\ll \bar{J}$ this is the appropriate limit to consider.  Taking into account this $T=0$ phase diagram leads to the conclusion that the azimuthal angle specifying intra-trimer interaction asymmetry must satisfy $\psi=\pi$ or $-\pi/4<\psi<\pi/4$.

In Fig.~\ref{fig5} the wave vector dependence of the energy integrated intensities at $T=10$ K are compared to the formulae of Eq.~(\ref{eq:B5}) and Eq.~(\ref{eq:B6}). The agreement between model and data is quite satisfactory considering that there are {\em no} adjustable parameters. While the comparison does not provide information on intra-trimer exchange, it supports the identification of a magnetic contribution to inelastic scattering at 7.5 meV, as the absolute cross section inferred from polarized and unpolarized magnetic neutron scattering is in perfect agreement with that predicted for the transition between the two doublets of the spin-trimer. 

The temperature dependence of the integrated intensities for the three inelastic peaks follows from Eq.~(\ref{eq:B5}) to Eq.~(\ref{eq:B7}). It depends only on the population of the intra-trimer levels, which in turn only depends on $E_{01}$ and $E_{02}$. The solid and dashed lines in Fig.~\ref{fig4}  (a) were calculated from these formulae using the experimental values for $E_{01}$ and $E_{02}$ and they are found to be in excellent agreement with the data.

\subsection{Magnetic Structure}
Only the ($\frac{1}{2}$00) magnetic Bragg peak could be detected in the present experiment. We obtained the absolute magnetic structure factor at $(\frac{1}{2}00)$ by comparison to the (230) nuclear Bragg peak. By combining this information with the symmetry analysis described in Appendix A, and the spectroscopic information presented in section~\ref{energylevels}, we can associate each of a few possible ordered structures with a specific set of intra-trimer exchange constants. 

The symmetry analysis is based on the assumption that any magnetic structure adopted through a second order phase transition can be expanded in basis functions for a single irreducible representation of the magnetic space group. Table~\ref{tab:basis} lists the basis functions for four of the irreducible representations of magnetically ordered \lcmo.  $\tau_1$ and $\tau_4$ describe uniaxial spin configurations with spins oriented along the {\bf c} direction while $\tau_2$ and $\tau_3$ correspond to co-planar structures with spins in the $\bf a-b$ plane.

\begin{table}
\caption{\label{tab:basis}The basis functions for irreducible representations of space group P112$_1$/m with magnetic wave vector ($\frac{1}{2}$00) and for atoms on the 2e site. The exact atomic coordinates are given in appendix A.}
\begin{ruledtabular}
\begin{tabular}{c|cccccc}
irreducible &\multicolumn{6}{c}{position (2e)} \\ \cline{2-7} 
representation&1&2&3&4&5&6 \\ \hline
$\tau_1$ & 001 & 000 & 000 & 00$\bar{1}$ & 000 & 000 \\
       & 000 & 001 & 000 & 000 & 00$\bar{1}$ & 000 \\
       & 000 & 000 & 001 & 000 & 000 & 00$\bar{1}$ \\ \hline
$\tau_2$ & 100 & 000 & 000 & 100 & 000 & 000 \\
       & 010 & 000 & 000 & 010 & 000 & 000 \\
       & 000 & 100 & 000 & 000 & 100 & 000 \\
       & 000 & 010 & 000 & 000 & 010 & 000 \\
       & 000 & 000 & 100 & 000 & 000 & 100 \\
       & 000 & 000 & 010 & 000 & 000 & 010 \\ \hline
$\tau_3$ & 100 & 000 & 000 & $\bar{1}$00 & 000 & 000 \\
       & 010 & 000 & 000 & 0$\bar{1}$0 & 000 & 000 \\
       & 000 & 100 & 000 & 000 & $\bar{1}$00 & 000 \\
       & 000 & 010 & 000 & 000 & 0$\bar{1}$0 & 000 \\
       & 000 & 000 & 100 & 000 & 000 & $\bar{1}$00 \\
       & 000 & 000 & 010 & 000 & 000 & 0$\bar{1}$0 \\ \hline
$\tau_4$ & 001 & 000 & 000 & 001 & 000 & 000 \\
       & 000 & 001 & 000 & 000 & 001 & 000 \\
       & 000 & 000 & 001 & 000 & 000 & 001 \\
\end{tabular}
\end{ruledtabular}
\end{table}

\begin{figure}
\includegraphics[width=3.2in]{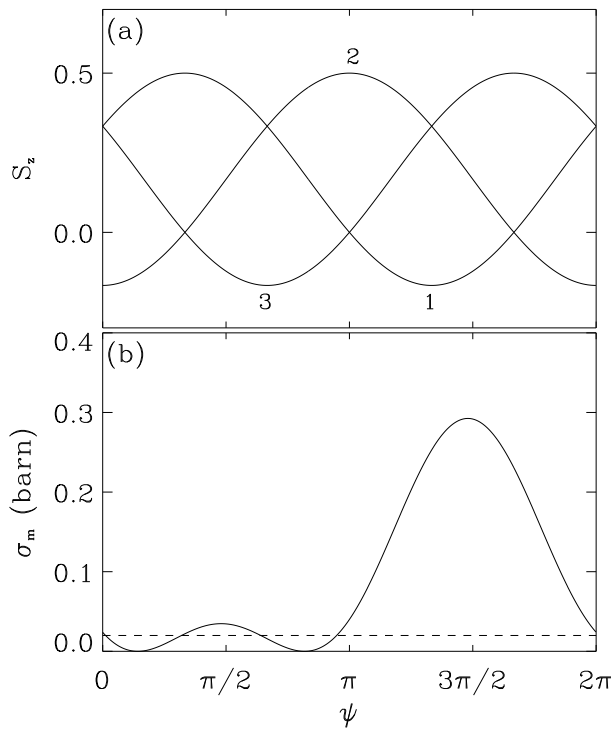}
\caption{\label{fig7}(a) Expectation value for the magnetic moment along a small applied field and at $T=0$ for  the three copper ions in a trimer as a function of index angle $\psi$. $\psi$ is the azimuthal angle spanning  intra-trimer exchange interactions that are consistent with a 7.5 meV doublet-doublet transition (see Fig.~\ref{fig6}). J$_1$/J$_3$ is equal to J$_2$/J$_3$ at $\psi=0$ and $\pi$. (b) Elastic magnetic neutron scattering cross section for the ($\frac{1}{2}$00) Bragg peak  per crystal unit cell calculated as a function of $\psi$ when the spins are along {\bf c} direction and adopt the magnetic structure associated with the $\tau_4$ irreducible representation. Dashed line shows the actual measured magnetic Bragg intensity.} 
\end{figure}

To make the connection between spectroscopy and magnetic structure we consider the ordered structure not as ordering of individual spins, but as ordering of the composite spin-1/2 degree of freedom associated with the lowest energy intra-trimer doublet. Appendix B lists the eigenstates of the intra-trimer spin Hamiltonian for arbitrary exchange interactions. N\'{e}el order corresponds to choosing a specific quantization axis on each spin trimer and alternating the doublet occupation consistent with the $(\frac{1}{2}00)$ magnetic wave-vector.  Fig.~\ref{fig7}(a) shows the spin projection on the quantization axis, for each of the three atoms on a trimer as a function of the azimuthal angle, $\psi$, that indexes possible intra-trimer exchange constants. Given a quantization axis, and a value for $\psi$, the magnetic structure factor for the $(\frac{1}{2}00)$ Bragg peak can be calculated for comparison with the measured absolute intensity of the magnetic Bragg peak. 

Spin structures corresponding to irreducible representations $\tau_1$ are inconsistent with the measured magnetic Bragg peak intensity for all values of $\psi$.  For $\tau_4$ Fig.~\ref{fig7}(b) shows that the calculated intensity is consistent with the  measured intensity, for four different values of $\psi=(0.01(2), 0.32(4), 0.64(4), 0.94(2))\pi$. Of these solutions, only the former two are consistent with the stability analysis for the $(\frac{1}{2}00)$ structure\cite{Wessel} and of these only the first is consistent with the approximately isosceles spin triangles. 

\begin{figure}
\includegraphics[width=3.2in]{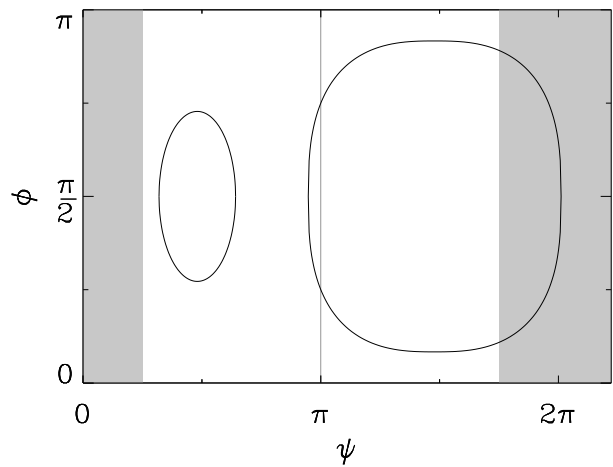}
\caption{\label{fig8}Combination of index angle $\psi$ and the spin orientation angle $\phi$ for magnetic structures with irreducible representation $\tau_2$ or $\tau_3$ that yields a ($\frac{1}{2}$00) magnetic Bragg intensity consistent with the experiment. Parameters in the hatched areas stabilize the ($\frac{1}{2}$00) magnetic structure for weak inter-trimer interactions\cite{Wessel}.} 
\end{figure}

Irreducible representations $\tau_2$ and $\tau_3$ describe spin configurations with spins in the {\bf a-b} plane. For simplicity we consider only uniaxial spin configurations spanned by the angle $\phi$ between the spin-trimer quantization axis and the $\bf a$ direction. $\tau_2$ and $\tau_3$ are different only in the registry of magnetic order in different $\bf a-b$ planes. It can be shown that $\tau_2$ and $\tau_3$ cannot be distinguished so they must be considered simultaneously. Fig.~\ref{fig8} shows values for $\psi$ and $\phi$ that are consistent with the measured $(\frac{1}{2}00)$ magnetic Bragg intensity. Also indicated as hatched areas are the values of $\psi$ where the $(\frac{1}{2}00)$ structure is stable.\cite{Wessel}

As previously mentioned the trimers have an approximate mirror plane, and this leads to the expectation that $J_1\approx J_2$. The seven spin structures and values of exchange constants that are consistent with this assumptions are listed in table~\ref{tab:spin}. Of these, the three structures with $\psi=0$ seem more likely because as opposed to the $\psi=\pi$ structures, they have a finite range of stability in the $J_1/J_3-J_2/J_3$ plane (see Fig.~\ref{fig6}).

\begin{table}
\caption{\label{tab:spin}Possible spin configurations corresponding to isosceles triangles with $J_1=J_2$. The exact atomic coordinates are given in appendix A.}
\begin{ruledtabular}
\begin{tabular}{c|ccccccccc}
&$\psi$ & $J_{1,2}$&$J_3$&{\bf S}$_1$&{\bf S}$_2$&{\bf S}$_3$&{\bf S}$_4$&{\bf S}$_5$&{\bf S}$_6$\\ 
&&meV&meV&&&&&&\\
\hline
$\tau_2$&0   &88.2&81.1& $0\frac{1}{3}0$ & $0\bar{\frac{1}{6}}0$ & $0\frac{1}{3}0$ & $0\frac{1}{3}0$ & $0\bar{\frac{1}{6}}0$ & $0\frac{1}{3}0$ \\ 
&$\pi$&83.2&91.1&000&$\frac{\sqrt{2}}{4}\frac{\sqrt{2}}{4}0$&000&000&$\frac{\sqrt{2}}{4}\frac{\sqrt{2}}{4}0$&000\\ 
&$\pi$&83.2&91.1&000&$\frac{\overline{\sqrt{2}}}{4}\frac{\sqrt{2}}{4}0$&000&000&$\frac{\overline{\sqrt{2}}}{4}\frac{\sqrt{2}}{4}0$&000\\ \hline
$\tau_3$& 0   &88.2&81.1& $0\frac{1}{3}0$ & $0\bar{\frac{1}{6}}0$ & $0\frac{1}{3}0$ & $0\bar{\frac{1}{3}}0$ & $0\frac{1}{6}0$ & $0\bar{\frac{1}{3}}0$ \\
&$\pi$&83.2&91.1&000&$\frac{\sqrt{2}}{4}\frac{\sqrt{2}}{4}0$&000&000&$\frac{\overline{\sqrt{2}}}{4}\frac{\overline{\sqrt{2}}}{4}0$&000\\
&$\pi$&83.2&91.1&000&$\frac{\overline{\sqrt{2}}}{4}\frac{\sqrt{2}}{4}0$&000&000&$\frac{\sqrt{2}}{4}\frac{\overline{\sqrt{2}}}{4}0$&000\\ \hline
$\tau_4$&0   &88.2&81.1& $00\frac{1}{3}$ & $00\bar{\frac{1}{6}}$ & $00\frac{1}{3}$ & $00\frac{1}{3}$ & $00\bar{\frac{1}{6}}$ & $00\frac{1}{3}$ \\
\end{tabular}
\end{ruledtabular}
\end{table}

\section{Conclusion}
In summary we have presented neutron scattering data that provide detailed microscopic information about frustrated quantum magnetism in  \lcmo . Neutron spectroscopy yields the average intra-trimer exchange constant of 85.8 meV and narrows possible intra-trimer exchange ratios to an elliptical trajectory in the $J_1/J_3-J_2/J_3$ plane. Neutron diffraction provides evidence for low temperature inter-trimer magnetic order that doubles the unit cell along the $\bf a$ direction. The $(\frac{1}{2}00)$ magnetic Bragg peak intensity and the approximate isosceles nature of the spin triangles narrows the possible exchange constants to $J_1=J_2=88.2$ meV and $J_3=81.1$ meV or $J_1=J_2=83.2$ meV and $J_3=91.1$ meV. A previously published mean field analysis of magnetic order in \lcmo\ indicates that the former combination of exchange constants is most likely. The corresponding ordered spin structures are uniaxial with two parallel spins at the base of the isosceles triangle  ($\langle S_z\rangle=\frac{1}{3}$) and an antiferromagnetically correlated spin of half the magnitude at the apex. The data are consistent with a spin direction either along $\bf b$ or $\bf c$. Another unresolved issue is the magnetic stacking sequence along $\bf c$ for which there are two options for spins oriented along $\bf b$.  

There is an instructive analogy between the frustrated cluster antiferromagnetism in \lcmo\ and rare earth magnets. In materials with Kramers rare earth ions, intra-atomic correlations establish effective spin-1/2 degrees of freedom, which subsequently develop long range magnetic order due to inter-atomic exchange interactions.\cite{mackintosh} Magnetic neutron scattering from such a system carries the rare earth atomic form factor.\cite{Lovesey} Magnetism in \lcmo\ is also based on a composite spin-1/2 degree of freedom, only it is spread over three atoms, it is established by inter-atomic exchange interactions, and it carries an oscillatory trimer ``form factor''. \lcmo\ is thus a particularly simple example of a concept of increasing importance in quantum magnetism. End states and holes in Haldane spin chains,\cite{afflecksorensen,xu}, impurity spins in high temperature superconductors,\cite{alloul} and spontaneously formed or structurally defined spin clusters in frustrated magnets.\cite{moessner,cheong} All are strongly correlated systems where suitably defined multi-atom composite spin degrees of freedom provide an enormous simplification for understanding low energy spin dynamics and the corresponding thermodynamic properties.

\begin{acknowledgments}
Work at JHU was supported by the NSF through DMR-0074571.
\end{acknowledgments}

\appendix
\section{Symmetry Analysis of Magnetic Ordering} 
A magnetic structure with a wave vector {\bf K} can be expanded in basis functions of a single irreducible representations of the space group of the crystal with wave vector {\bf K}:
\begin{equation}
\label{eq:A1}
{\bf S}_{0j}^{{\bf K}\nu}=\sum_\lambda C_\lambda^\nu {\bf S}_0(_\lambda^{{\bf K}\nu}|j)~,
\end{equation}
where ${\bf S}_{0j}^{{\bf K}\nu}$ describes the spin vector of the jth magnetic ion in the 0th cell which determines the magnetic structure of the crystal with a wave vector {\bf K}, and ${\bf S}_0(_\lambda^{{\bf K}\nu}|j)$ is the basis function transforming according to the $\nu$th irreducible representation. The spin vector in the nth cell of the crystal can be derived from the spins in the zeroth cell by the equation
\begin{equation}
\label{eq:A2}
{\bf S}_{n}(^{{\bf K}\nu}_\lambda |j)=e^{i{\bf K}\cdot{\bf t}_n} {\bf S}_0(_\lambda^{{\bf K}\nu}|j)~.
\end{equation}

In accordance with the Landau theory of second order phase transitions, the majority of magnetic structures are characterized by a single irreducible representation. However, there are cases that involve two or more irreducible representations of the magnetic space group. We will follow the method developed by Izyumov and Naish\cite{Izyumov} to calculate the basis functions, considering only magnetic structures defined through a single irreducible representation.
 
The magnetic representation $d_M^{\bf K}$ may be expanded into irreducible representations $d^{{\bf K}\nu}$ of the wave vector group $G_{\bf K}$ by
\begin{eqnarray}
d_M^{\bf K}&=&\sum_\nu n_\nu d^{{\bf K}\nu}~, \label{eq:A3}\\
n_\nu&=&\frac{1}{n(G_{\bf K}^0)}\sum_{h\in G_{\bf K}^0}\chi_M^{\bf K}(g)\chi^{*{\bf K}\nu}(g)~,\label{eq:A4}\\
\chi_M^{\bf K}(g)&=&\delta_hSpR^h\sum_je^{-i{\bf K}\cdot{\bf a}_p(g,j)}\delta_{j,gj}~, \label{eq:A5}
\end{eqnarray}
where $\chi^{{\bf K}\nu}(g)$ is the character of irreducible representation $d^{{\bf K}\nu}$ of group $G_{\bf K}$ and $\chi_M^{\bf K}$ is the character of the magnetic representation. $\delta_h$ equals 1 for usual rotations and -1 for inversion rotations. $SpR^h$ is the trace of the matrix $R_{\alpha\beta}^h$ of rotation for the group element g, which includes a rotation part h and a translation part ${\bf \tau}_h$:
\begin{equation}
\label{eq:A6}
g{\bf x}_j=h{\bf x}_j+{\bf \tau}_h={\bf x}_i+{\bf a}_p(g,j)~.
\end{equation}
The summation in equation (\ref{eq:A4}) is taken only over the zero block of group G$_{\bf K}$. The magnetic atomic components of the basis functions of the irreducible representation for the 0th cell are
\begin{equation}
\label{eq:A7}
{\bf S}_0(_\lambda^{{\bf K}\nu}|i)=\sum_{h\in G_{\bf K}^0}\delta_h d_{\lambda[\mu]}^{*{\bf K}\nu}(g)e^{-i{\bf K}\cdot {\bf a}_p(g,j)}\delta_{i,g[j]} \left(\begin{array}{l} R_{x[\beta]}^h \\R_{y[\beta]}^h \\R_{z[\beta]}^h \end{array} \right) ~.
\end{equation}

The space group of \lcmo\ is P112$_1$/m. There are six magnetic copper ions in the primitive cell occupying the positions
\begin{eqnarray*}
2(e)&:&1(x_1,~y_1,~0),~2(x_2,~y_2,~0),~3(x_3,~y_3,~0),\\
&&4(1-x_1,~1-y_1,~0.5),~5(1-x_2,~1-y_2,~0.5),\\
&&6(1-x_3,~1-y_3,~0.5),\\
\end{eqnarray*}
where ($x_i$, $y_i$) are  (1.0893, 0.8926), (0.8782, 0.4623), and 
(0.6465, 0.8816) for i=1, 2, and 3 respectively.
The coordinates are written in the Kovalev system\cite{Kovalev}, and are shifted by (00$\bar{\frac{1}{4}}$) from the coordinates listed in the International Table of Crystallography.\cite{xraytables}

Group P112$_1$/m contains four elements, which are listed in table~\ref{tab:ele}. Table \ref{tab:ele} also shows the permutations of atoms by the action of the group elements. 
\begin{table}
\caption{\label{tab:ele}Permutation of Cu atoms in \lcmo\ crystal by the elements of group P112$_1$/m. ${\bf a}_p$ is the returning translation vector.}
\begin{ruledtabular}
\begin{tabular}{lcl}
element&atoms&${\bf a}_p$ \\ \cline{2-2}
 &1 2 3 4 5 6& \\ \hline
\{h$_1|$000\} & 1 2 3 4 5 6&(000) \\
\{h$_4|$00$\frac{1}{2}$\} & 1 2 3 4 5 6&(110)$_{1-3}$, (11$\bar{1}$)$_{4-6}$\\
\{h$_{25}|$00$\frac{1}{2}$\} & 4 5 6 1 2 3&(110) \\
\{h$_{28}|$000\} & 1 2 3 4 5 6&(000)$_{1-3}$, (001)$_{4-6}$\\
\end{tabular}
\end{ruledtabular}
\end{table}
The irreducible representations of the wave vector group\cite{Kovalev} are listed in table~\ref{tab:irre}. The magnetic representation for the wave vector {\bf K}=($\frac{1}{2}$00) can be decomposed into irreducible representations as follows:
\begin{equation}
\label{eq:A8}
d_M^{\bf K}=3\tau_1+6\tau_2+6\tau_3+3\tau_4 ~.
\end{equation}

\begin{table}
\caption{\label{tab:irre}Irreducible representations of group $C_{2h}^2$ for {\bf K}=$\frac{1}{2}{\bf a}^*$.}
\begin{ruledtabular}
\begin{tabular}{c|cccc}
T4 & h$_1$ & h$_4$ & h$_{25}$ & h$_{28}$ \\ \hline
$\tau_1$ & 1 & 1 & 1 & 1 \\
$\tau_2$ & 1 & 1 &-1 &-1 \\
$\tau_3$ & 1 &-1 & 1 &-1 \\
$\tau_4$ & 1 &-1 &-1 & 1 \\ 
\end{tabular}
\end{ruledtabular}
\end{table}

The calculated basis functions are shown in table~\ref{tab:basis}. Since the actual spin directions can be combinations of the basis functions within a single irreducible representation, it follows from the table that the spins have to be  either along {\bf c} direction or in the {\bf a-b} plane. For the irreducible representations $\tau_1$ and $\tau_3$, the spins on sites 4-6 are antiparallel to those on sites 1-3, while they are parallel for the irreducible representations $\tau_2$ and $\tau_4$.

\section{Neutron Scattering from a Trimer}
The  Hamiltonian for a spin triangle with Heisenberg exchange interactions is
\begin{equation}
\label{apeq:1}
H=J_1 {\bf S_1 \cdot S_2}+J_2 {\bf S_2 \cdot S_3} +J_3 {\bf S_3 \cdot S_1}~. 
\end{equation}
The eigenstates and eigenvalues are listed in table \ref{tab:eigens}.
\begin{table}
\caption{\label{tab:eigens}Eigenvalues and eigenstates of the Hamiltonian H. The eigenstates listed in the table are not normalized. We use the following abbreviations: $J=\frac{1}{3}(J_1+J_2+J_3)$, 
$a=J_1J_2-J_3^2-J_3\Delta$, 
$b=J_1J_3-J_2^2-J_2\Delta$, 
$c=J_2J_3-J_1^2-J_1\Delta$, 
$A=J_1J_2-J_3^2+J_3\Delta$, 
$B=J_1J_3-J_2^2+J_2\Delta$, 
$C=J_2J_3-J_1^2+J_1\Delta$, and 
$\Delta=\sqrt{J_1^2+J_2^2+J_3^2-J_1J_2-J_2J_3-J_3J_1}$}
\begin{ruledtabular}
\begin{tabular}{ccc}
&E &$|\phi\rangle$ \\ \hline\\
0&$(-3J-2\Delta)/4$ &
    $-\frac{a+c}{a}\alpha \beta \beta+\frac{c}{a} \beta \alpha \beta+\beta \beta \alpha$  \\ \\
& & $-\frac{a+b}{a} \alpha\alpha\beta +\frac{b}{a} \alpha\beta\alpha+\beta\alpha\alpha$ \\ \\ \hline \\
1&$(-3J+2\Delta)/4$ & 
    $-\frac{A+C}{A}\alpha \beta \beta+\frac{C}{A} \beta \alpha \beta+\beta \beta \alpha$  \\ \\
 &&$-\frac{A+B}{A} \alpha\alpha\beta +\frac{B}{A} \alpha\beta\alpha+\beta\alpha\alpha$ \\  \\ \hline \\
2&$3J/4$ & $\alpha\alpha\alpha$\\ \\
 && $\beta\beta\beta$\\ \\
 && $\alpha\beta\beta+\beta\alpha\beta+\beta\beta\alpha$\\ \\
 && $\alpha\alpha\beta+\alpha\beta\alpha+\beta\alpha\alpha$\\ \\ 
\end{tabular}
\end{ruledtabular}
\end{table}
The differential magnetic neutron cross section for inelastic transitions $|S\rangle \rightarrow |S'\rangle $ is\cite{Lovesey}
\begin{eqnarray}
\frac{d^2\sigma}{d\Omega d\omega}&= &
C_0 \rho(S)\sum_{\alpha\beta}\left(\delta_{\alpha\beta}-\hat{Q}_\alpha\hat{Q_\beta} \right)\times\nonumber\\
&&\sum_{jj'}exp\left[i {\bf Q}\cdot({\bf R}_j-{\bf R}_{j'})\right] \nonumber \\
&&\sum_{MM'}\langle SM|\hat{S}^\alpha_j |S'M'\rangle \langle S'M'|\hat{S}^\beta_{j'} |SM\rangle \times\nonumber\\ 
&&\delta\left(\hbar\omega+E(S)-E(S')\right)~, \label{eq:B1}
\end{eqnarray}
where
\begin{eqnarray*}
C_0&=&N\left(\frac{\gamma e^2}{m_e c^2} \right)\frac{k'}{k}F^2({\bf Q})exp\left[-2 W({\bf Q})\right]~, \\
\rho(S)&=&Z^{-1}\exp\left[-\frac{E(S)}{k_BT}\right]~.\\
\end{eqnarray*} 
Also, ${\bf Q}={\bf k}-{\bf k'}$ is the scattering wave vector, $F({\bf Q})$ is the magnetic form factor,\cite{xraytables} and Z is the partition function. It can be shown that terms in the cross section with $\alpha\neq\beta$ vanish. After averaging over all directions for ${\bf Q}$ we obtain the following cross section for a powder sample
\begin{eqnarray}
\left(\frac{d^2\sigma}{d\Omega d\omega} \right)_{0\rightarrow 1}&=&C_0\rho(0)\frac{2}{3}\left[ 1+\right.\nonumber\\
&&\frac{sin(QR_{12})}{QR_{12}}\frac{(J_1-J_2)(J_3-J_1)}{\Delta^2}+\nonumber\\   &&\frac{sin(QR_{23})}{QR_{23}}\frac{(J_1-J_2)(J_2-J_3)}{\Delta^2}+\nonumber\\
&&\left.\frac{sin(QR_{13})}{QR_{13}}\frac{(J_1-J_3)(J_3-J2)}{\Delta^2}\right]\times\nonumber \\
&&\delta(\hbar\omega+E_0-E_1)~,\label{eq:B2}\\
\left(\frac{d^2\sigma}{d\Omega d\omega} \right)_{0\rightarrow 2}&=&C_0 \rho(0)\frac{2}{3}\left[ 2+\frac{sin(QR_{12})}{QR_{12}}\right.\times\nonumber\\
&&\left(\frac{-a(a+b)}{a^2+b^2+ab}+\frac{-bc}{b^2+c^2-bc} \right)+\nonumber\\
&&\frac{sin(QR_{13})}{QR_{13}}\left(\frac{ab}{a^2+b^2+ab}+\right.\nonumber\\
&&\left.\frac{b(c-b)}{b^2+c^2-bc} \right)+\frac{sin(QR_{23})}{QR_{23}}\times\nonumber\\
&&\left.\left(\frac{-b(a+b)}{a^2+b^2+ab}+\frac{-c(c-b)}{b^2+c^2-bc} \right) \right]\times\nonumber\\
&&\delta(\hbar\omega+E_0-E_2)~,\label{eq:B3}\\
\left(\frac{d^2\sigma}{d\Omega d\omega} \right)_{1\rightarrow 2}&=&C_0 \rho(1)\frac{2}{3}\left[ 2+\frac{sin(QR_{12})}{QR_{12}}\right.\times\nonumber\\
&&\left(\frac{-A(A+B)}{A^2+B^2+AB}+\frac{-BC}{B^2+C^2-BC} \right)+\nonumber\\
&&\frac{sin(QR_{13})}{QR_{13}}\left(\frac{AB}{A^2+B^2+AB}+\right.\nonumber\\
&&\left.\frac{B(C-B)}{B^2+C^2-BC} \right)+\frac{sin(QR_{23})}{QR_{23}}\times\nonumber\\
&&\left(\frac{-B(A+B)}{A^2+B^2+AB}+\right.\nonumber\\
&&\left.\left.\frac{-C(C-B)}{B^2+C^2-BC} \right) \right]\times\nonumber\\
&&\delta(\hbar\omega+E_1-E_2)~.\label{eq:B4}
\end{eqnarray}
The spin triangles in \lcmo\  are close to equilateral ($R_{12}=3.39$ ${\rm \AA}$, $R_{13}=3.50$ ${\rm \AA}$, $R_{23}=3.41$ ${\rm \AA}$), so we can replace the Cu-Cu distances by R=$\frac{1}{3}(R_{12}+R_{23}+R_{13})$:
\begin{eqnarray}
\left(\frac{d^2\sigma}{d\Omega d\omega} \right)_{0\rightarrow 1}&=&C_0\rho(0)\frac{2}{3}\left[1-\frac{sin(QR)}{QR}\right]\times\nonumber\\
&&\delta(\hbar\omega+E_0-E_1)~,\label{eq:B5}\\
\left(\frac{d^2\sigma}{d\Omega d\omega} \right)_{0\rightarrow 2}&=&C_0\rho(0)\frac{4}{3}\left[1-\frac{sin(QR)}{QR}\right]\times\nonumber\\
&&\delta(\hbar\omega+E_0-E_2)~,\label{eq:B6}\\
\left(\frac{d^2\sigma}{d\Omega d\omega} \right)_{1\rightarrow 2}&=&C_0\rho(1)\frac{4}{3}\left[1-\frac{sin(QR)}{QR}\right]\times\nonumber\\
&&\delta(\hbar\omega+E_1-E_2)~.\label{eq:B7}
\end{eqnarray}
As each level is degenerate there is also an elastic cross section associated with intra-level transitions. However, this is not relevant for the experiment and so will not be listed.


\end{document}